\documentclass[a4paper, oneside, 10pt]{article}
\usepackage{eurosym}
\usepackage{libertine}
\usepackage[english]{babel}
\usepackage[utf8]{inputenc}
\usepackage{enumerate}
\usepackage{graphicx}
\usepackage[hidelinks]{hyperref}
\usepackage{fancyhdr}
\usepackage{amsthm}
\usepackage{amssymb}
\usepackage{amsmath}
\usepackage{amsfonts}
\usepackage{minibox}
\usepackage{float}
\usepackage{graphicx}
\usepackage{subcaption}
\usepackage[export]{adjustbox}
\usepackage{wrapfig}
\usepackage[strict]{changepage}
\usepackage{listings}
\usepackage{listingsutf8}
\usepackage{color}
\usepackage{tikz}
\usetikzlibrary{calc}
\usepackage{pgfplots}
\pgfplotsset{width=13cm,compat=1.9}
\usepgfplotslibrary{external}
\usepackage{framed}
\usepackage{braket}
\usepackage{longtable}
\usepackage{epigraph}
\usepackage[numbers,sort&compress]{natbib}
\usepackage{ragged2e}
\usepackage{caption}

\numberwithin{equation}{section}
\input xypic
\theoremstyle{remark}

\DeclareMathOperator{\Lag}{\mathcal{L}}
\date{}
\title{\textbf{Cosmological Reconstruction in $f(Q, T)$ Gravity: $\Lambda$CDM Background and Matter-Sector Dependence}}
\author{Shivank Pal$^{1}$ \thanks{Email: shivankpal1997@gmail.com}, Gauree Shanker$^{1}$\thanks{Email:  gauree.shanker@cup.edu.in}
 \vspace{0.3cm}\\
 ${}^{1}$ Department of Mathematics and Statistics,\\Central University of Punjab, Bathinda, 151401, Punjab, India. 
 }

\begin{document}
\maketitle
\begin{abstract}
In this work, we investigate the cosmological reconstruction of $f(Q, T)$ gravity, where $Q$ is the non-metricity scalar and $T$ is the trace of the energy-momentum tensor. We consider the functional form $f(Q,T)=f(Q)+\lambda T$ and reconstruct explicit forms of $f(Q)$ corresponding to the $\Lambda$CDM expansion history in Friedmann-Lema\^itre-Robertson-Walker (FLRW) universe. By employing the matter conservation equation and expressing the cosmological quantities in terms of $Q$, the reconstruction problem is formulated as a first-order linear differential equation for $f(Q)$. Analytical solutions for $f(Q)$ are obtained for various matter configurations, including dust-like matter, a perfect fluid with equation-of-state parameter $\omega=-1/3$, and a nonisentropic perfect fluid with a time-dependent barotropic index. Additionally, an e-folding formulation of the reconstruction is considered to examine the cosmological evolution in terms of the e-folding parameter. The resulting forms of $f(Q)$ confirm that the $\Lambda$CDM expansion history can be successfully realized within the $f(Q,T)$ framework across diverse matter sectors.
\end{abstract}

\section{Introduction}
The current acceleration of the expansion of the universe has been a major problem for the past two decades. Recent developments in observational cosmology, including Type Ia supernovae \cite{riess1998observational, perlmutter1999measurements}, Cosmic Microwave Background radiation \cite{spergel2003first}, and large-scale structure observations \cite{nadathur2020testing,daniel2008large,koivisto2006dark}, provide strong evidence for the accelerated expansion of the universe. The general theory of relativity has undoubtedly been the most successful theory of gravity in describing the large-scale structure of the universe. In cosmology, the Friedmann-Lema\^itre-Robertson-Walker (FLRW) metric, together with a suitable matter content, reduces the gravitational field equations to those governing the scale factor $a(t)$, enabling the study of the expansion history of the universe. A possible explanation for the current accelerated expansion of the universe is the addition of a cosmological constant in the Einstein field equations. This is the standard model of cosmology, known as the $\Lambda$CDM (Lambda Cold Dark Matter) model, which can explain the late-time cosmic acceleration of the universe and fit all the available observations discussed previously. But still, this model has problems, such as fine-tuning issues and the coincidence problem \cite{di2021realm,carroll2001cosmological,padmanabhan2003cosmological}. Thus, it is crucial to investigate alternative approaches to gravity theories.\\
Typically, there are two ways to address these issues: by modifying the matter sector by including an additional dark component in the universe's energy content, or by modifying general relativity. In addition to the conventional curvature-based representation, General Relativity (GR) can be formulated in terms of torsion or nonmetricity. The torsion-based formulation is known as the Teleparallel Equivalent of GR (TEGR), whereas the nonmetricity-based formulation is known as the Symmetric Teleparallel Equivalent of GR (STEGR). In TEGR, a curvature-free and metric-compatible geometry is used, and the gravity is described through torsion, while STEGR is formulated in a flat and torsion-free geometry where non-metricity provides the geometrical description of gravity. $f(T)$ \cite{cai2016f} and $f(Q)$ \cite{jimenez2018coincident} theories are the corresponding modifications of these theories, respectively. As a viable framework for studying the late-time accelerated expansion of the universe, $f(Q)$ gravity has attracted significant attention in cosmology \cite{heisenberg2024review}. Moreover, $f(Q)$ gravity has been considered to explain the late-time acceleration and dark energy concerns \cite{jimenez2020cosmology,atayde2021can}.
By introducing a coupling between the non-metricity scalar $Q$ and the trace of the energy-momentum tensor $T$, the $f(Q)$ framework can be extended, and this leads to the generalized $f(Q, T)$ theory \cite{xu2019f}. Here, the gravitational action is described by a general function of $Q$ and $T$ that allows the geometry and matter sectors to be coupled in the gravitational Lagrangian. Several studies have been done in $f(Q, T)$ gravity by considering specific and relatively simple functional forms of $f(Q, T)$, which demonstrate that it can account for the observed late-time cosmic acceleration and provide an alternative to the dark energy problem \cite{arora2021constraining,gadbail2022generalized}. However, rather than prescribing the complete functional form of $f(Q, T)$, one may instead reconstruct the unknown Q-dependent part $f(Q)$ from a prescribed cosmological expansion history.\\

\noindent On the other hand, to investigate modified theories of gravity by determining the gravitational Lagrangian compatible with a prescribed cosmological expansion history, the cosmological reconstruction approach provides a systematic framework. In contrast to approaches that begin with a fully specified gravitational function and subsequently study its cosmological evolution, reconstruction starts from a desired background evolution and employs the corresponding field equations to determine the unknown functional dependence of the gravitational Lagrangian. In practical applications, an appropriate ansatz can be adopted for part of the gravitational function, while the remaining functional dependence is reconstructed from the prescribed cosmological evolution. This approach is particularly useful in modified gravity, where the complexity and nonlinearity of the field equations can make it difficult to obtain exact solutions and establish a direct connection between theoretical models and cosmological observations.\\
Cosmological reconstruction has been extensively studied in $f(R)$ gravity across various scenarios, particularly to obtain viable cosmological evolutions that connect the matter-dominated era to the late-time dark-energy-dominated phase. Similar techniques have subsequently been applied to other modified gravity frameworks \cite{nojiri2006modified, capozziello2006cosmological, nojiri2009cosmological, dunsby2010lambda}. For instance, reconstruction methods in $f(Q)$ gravity have been employed to investigate isotropic and anisotropic cosmological solutions \cite{esposito2022reconstructing}. Furthermore, related reconstruction approaches have been used to obtain cosmological evolutions characterized by power-law expansion, de Sitter solutions, and phantom or non-phantom phases in various extended theories of gravity \cite{elizalde2010lambdacdm,houndjo2012reconstructing,jamil2012reconstruction}. These studies demonstrate the usefulness of the reconstruction framework in connecting a prescribed cosmological history with the underlying gravitational theory.

\noindent Although the reconstruction of $f(Q,T)$ Lagrangians across various cosmological scenarios was investigated in Reference \cite{gadbail2023reconstruction}, their study relies on several functional ansatzes for $f(Q,T)$ to examine a broad range of cosmic evolutionary behaviors. In contrast, we present a systematic reconstruction of the gravitational Lagrangian specifically required to reproduce a background $\Lambda$CDM expansion history within the functional framework $f(Q,T) = f(Q) + \lambda T$. We derive exact analytical solutions for dust-like matter, a perfect fluid with $\omega = -1/3$, and a nonisentropic perfect fluid, followed by an alternative formulation in terms of the e-folding parameter.

\noindent In this work, we investigate the cosmological reconstruction of the gravitational Lagrangian within the $f(Q,T)$ framework by considering the functional form $f(Q,T)=f(Q)+\lambda T$, where $\lambda$ is the matter--geometry coupling parameter. We assume the cosmological expansion history to follow the standard $\Lambda$CDM background and reconstruct the unknown function $f(Q)$ corresponding to this prescribed evolution. Using the modified Friedmann equation, together with the relation $Q=6H^2$ and the trace relation $T=-\rho+3p$, we formulate the reconstruction problem as a first-order linear ordinary differential equation for $f(Q)$. The matter density is then expressed in terms of the non-metricity scalar for different matter configurations, and the corresponding reconstructed forms of $f(Q)$ are obtained. In particular, we consider a dust-like matter source with $\omega=0$, a perfect fluid with $\omega=-1/3$, and a nonisentropic perfect fluid characterized by a time-dependent barotropic index. Furthermore, we reformulate the reconstruction procedure in terms of the e-folding parameter to investigate the cosmological evolution from an alternative parametrization. The resulting reconstructed gravitational Lagrangians demonstrate how the prescribed $\Lambda$CDM expansion history can be realized within the considered $f(Q,T)$ framework for different choices of the matter sector.\\

\noindent The remainder of this paper is organized as follows. In Section (\ref{2}), we present the $f(Q,T)$ gravity formalism and the corresponding cosmological field equations. In Section (\ref{3}), we reconstruct the gravitational Lagrangian within the $f(Q,T)$ framework for the prescribed $\Lambda$CDM expansion history. In particular, Subection (\ref{3.1}) considers the reconstruction for dust-like matter, Subection (\ref{3.2}) deals with a perfect fluid with $\omega=-1/3$, and Subection (\ref{3.3}) is devoted to the reconstruction for nonisentropic perfect fluids. In Section (\ref{4}), the reconstruction procedure is reformulated in terms of the e-folding parameter. Finally, the main conclusions of the work are summarized in Section (\ref{5}).

\section{$f(Q, T)$ Formalism} \label{2}
For $f(Q, T)$ gravity, the action is given as \cite{xu2019f}
\begin{equation*}
    S = \int \left(\frac{1}{16 \pi} f(Q, T) + \Lag_m \right) \sqrt{-g}~d^4 x 
    \tag{1} \label{equ(1)}
\end{equation*}
where $f$ is an arbitrary function of the non-metricity scalar $Q$ and $T$ is the trace of the energy-momentum tensor, $\Lag_m$ represents the matter Lagrangian and $g = det (g_{\mu \nu})$. The non-metricity scalar $Q$ is defined as
\begin{equation*}
    Q = - g^{\mu \nu} (L^{\alpha}_{\beta \mu} L^{\beta}_{\nu \alpha} - L^{\alpha}_{\beta \alpha} L^{\beta}_{\mu \nu})
    \tag{2} \label{equ(2)}
\end{equation*}
where $L^{\alpha}_{\beta \gamma}$ is the deformation tensor given by
\begin{equation*}
    L^{\alpha}_{\beta \gamma} = - \frac{1}{2}g^{\alpha \lambda}(\nabla_{\gamma}g_{\beta \lambda} + \nabla_{\beta}g_{\lambda \gamma} - \nabla_{\lambda} g_{\beta \gamma})
    \tag{3} \label{equ(3)}
\end{equation*}
The non-metricity tensor is defined as $Q_{\alpha \mu \nu} \equiv - \nabla_{\alpha} g_{\mu \nu}$, with its trace given by $Q_{\alpha} = Q_{\alpha \mu}^{\mu}$. The energy-momentum tensor and its trace are defined as
\begin{equation*}
     T_{\mu \nu} = - \frac{2}{\sqrt{-g}} \frac{\delta(\sqrt{-g} \Lag_M)}{\delta g^{\mu \nu}}, \quad T = g^{\mu \nu}T_{\mu \nu}.
    \tag{4} \label{equ(4)}
\end{equation*}
And $\Theta_{\mu \nu}$ is defined as
\begin{equation*}
    \Theta_{\mu \nu} = g^{\alpha \beta} \frac{\delta T_{\alpha \beta}}{\delta g^{\mu \nu}}.
    \tag{5} \label{equ(5)}
\end{equation*}
The variation of the gravitational action leads to the following field equation
\begin{equation*}
    8 \pi T_{\mu \nu} = - \frac{2}{\sqrt{-g}} \nabla_{\alpha}\left(f_Q \sqrt{-g} P^{\alpha}_{\mu \nu}\right) - \frac{1}{2}f g_{\mu \nu} + f_T (T_{\mu \nu} + \Theta_{\mu \nu}) - f_Q (P_{\mu \alpha \beta} Q^{\alpha \beta}_{\nu} - 2 Q^{\alpha \beta}_{\mu} P_{\alpha \beta \nu}) 
    \tag{6} \label{equ(6)}
\end{equation*}
where $f_Q = \frac{\partial f(Q, T)}{\partial Q},~ f_T = \frac{\partial f(Q, T)}{\partial T}$ and $P^{\alpha}_{\mu \nu}$, is the superpotential of the model, given as
\begin{equation*}
    P^{\alpha}_{\mu \nu} = - \frac{1}{2}L^{\alpha}_{\mu \nu} + \frac{1}{4} \left(Q^{\alpha} - \Tilde{Q}^{\alpha} \right) g_{\mu \nu} - \frac{1}{4} \delta^{\alpha}_{(\mu}Q_{\nu )},
    \tag{7} \label{equ(7)}
\end{equation*}
We now assume a flat FLRW metric as,
\begin{equation*}
    ds^2 = -N^2(t)dt^2 + a(t)^2 (dx^2 + dy^2 + dz^2)
    \tag{8} \label{equ(8)}
\end{equation*}
where $a(t)$ is the scale factor and $N(t)$ is a Lapse function. We adopt the coincident gauge $N(t) = 1$, in which the covariant derivatives reduce to ordinary derivatives \cite{jimenez2018coincident,jimenez2020cosmology}. For a homogeneous and isotropic FLRW spacetime with scale factor $a(t)$, the Hubble parameter is defined as $H = \frac{\dot{a}}{a}$ and the non-metricity scalar reduces to $Q = 6H^2$. The trace of the energy-momentum tensor for a perfect fluid is given as $T = -\rho +3p$ where $\rho$ and $p$ denote the energy density and pressure. The generalized Friedman equations of $f(Q, T)$ theory are,
\begin{equation*}
    8 \pi \rho = \frac{f}{2} - 6FH^2 - \frac{2 \Tilde{G}}{1+ \Tilde{G}}(\dot F H + F \dot H),
    \tag{9} \label{equ(9)}
\end{equation*}
and 
\begin{equation*}
    8 \pi p = - \frac{f}{2} + 6FH^2 + 2(\dot F H + F \dot H),
    \tag{10} \label{equ(10)}
\end{equation*}
where dot represents derivative with respect to time and $F = f_Q$ and $8 \pi \Tilde{G} = f_T$.
Using the above equations \eqref{equ(9)} and \eqref{equ(10)}, we can write the equations in a form similar to that of standard general relativity(GR),
\begin{equation*}
    3H^2 = 8\pi \rho_{eff} = \frac{f}{4F} - \frac{4 \pi}{F} [(1 + \Tilde{G}) \rho + \Tilde{G} p],
    \tag{11} \label{equ(11)}
\end{equation*}
and 
\begin{equation*}
    2 \dot H + 3 H^2 = - 8 \pi p_{eff} = \frac{f}{4F} - \frac{2 \dot F H}{F} + \frac{4 \pi}{F}[(1+ \Tilde{G})\rho +(2 + \Tilde{G})p].
    \tag{12} \label{equ(12)}
\end{equation*}
Moreover, $\rho_{eff}$ and $p_{eff}$ are the effective density and effective pressure, respectively.\\
Now, one can investigate various applications using the above Friedmann equations in the background of $f(Q, T)$ gravity. Further, the conservation equation is \cite{xu2019f}
\begin{equation*}
    \dot \rho_{eff} + 3H (\rho_{eff}+ p_{eff}) = 0
    \tag{13} \label{equ(13)}
\end{equation*}
where $F \equiv f_Q$ and $8 \pi \Tilde{G} \equiv f_T$

\section{Reconstruction of the Gravitational Lagrangian in \texorpdfstring{$f(Q,T)$}{f(Q,T)} Gravity} \label{3}

In this section, we reconstruct the gravitational Lagrangian corresponding to the $\Lambda$CDM cosmological model within the framework of modified symmetric teleparallel gravity. We consider the following functional form of $f(Q, T)$ 
\begin{equation*}
    f(Q,T)=f(Q)+\lambda T,
    \tag{14} \label{equ(14)}
\end{equation*}
where $\lambda$ is the matter--geometry coupling parameter. In the upcoming subsections, we aim to carry out the reconstruction procedure, which consists of determining the explicit form of the function $f(Q)$ that reproduces the prescribed cosmological evolution. The reconstruction is carried out for different matter sources in order to investigate the influence of the equation of state on the resulting gravitational Lagrangian.\\
We assume that the matter sector is described by a perfect fluid that obeys the equation of state $p=\omega\rho,$ together with the trace relation $ T=-\rho+3p,$ the modified Friedmann equation \eqref{equ(11)} can be rewritten as
\begin{equation*}
 3H^2 = \frac{f(Q) + \lambda T}{4f_Q} - \frac{4\pi}{f_Q} \left[1 + \frac{\lambda (1+ \omega)}{8 \pi} \right] \rho.
 \tag{15} \label{equ(15)}
\end{equation*}
Finally, using the relation $Q=6H^2$, we obtain the first-order linear ordinary differential equation for the unknown function $f(Q)$
\begin{equation*}
  2Qf_Q - f(Q) = - \rho \left[16 \pi + \lambda (3 - \omega) \right].
  \tag{16} \label{equ(16)}
\end{equation*}
To reconstruct the explicit form of $f(Q)$, we now specify the cosmological expansion history. Throughout this paper, we assume that the expansion history follows the standard $\Lambda$CDM cosmology, whose Hubble parameter is given as
\begin{equation*}
    H^2(z) = H_0^2 [\Omega_{m_0} (1+z)^3 + \Omega_{\Lambda}],
    \tag{17} \label{equ(17)}
\end{equation*}
where $\Omega_{m_0}$ is the matter--density parameter and $\Omega_{\Lambda}$ dark energy (cosmological constant) density parameter. Here, the $\Lambda$CDM expansion history is treated as the prescribed background for the reconstruction, while different matter sources are subsequently considered within this background. Using the relation between the scale factor and the cosmological redshift $a=\frac{1}{1+z},$ equation \eqref{equ(17)} can be rewritten in terms of the scale factor,
\begin{equation*}
H^2(a) = H_0^2 \left(\frac{\Omega_{m_0}}{a^3} + \Omega_{\Lambda} \right).
\tag{18} \label{equ(18)}
\end{equation*}
Employing the definition of the non-metricity scalar $Q=6H^2,$ we obtain the non-metricity scalar as a function of the scale factor,
\begin{equation*}
    Q(a) = 6H_0^2 \left(\frac{\Omega_{m_0}}{a^3} + \Omega_{\Lambda}\right)
    \tag{19} \label{equ(19)}
\end{equation*}
which can be inverted to express the scale factor in terms of the non-metricity scalar,

\begin{equation*}
    a(Q) = \left(\frac{6H_0^2 \Omega_{m_0}}{Q - 6H_0^2 \Omega_{\Lambda}} \right)^{\frac{1}{3}}
    \tag{20} \label{equ(20)}
\end{equation*}
The above relation provides the key ingredient required for the reconstruction procedure. The corresponding energy density, for each matter source considered below, is first obtained in terms of the scale factor from the conservation equation. This density can be expressed entirely in terms of the non-metricity scalar $Q$ using the equation \eqref{equ(20)}. Substituting into the differential equation \eqref{equ(16)} yields a first-order linear differential equation for $f(Q)$, which can then be solved to obtain the reconstructed gravitational Lagrangian.

\subsection{Reconstruction for Dust-Like Matter \texorpdfstring{$\omega = 0$}{w = 0}} \label{3.1}
We first consider the pressureless matter-dominated universe, for which the equation of state parameter is $\omega=0$. In this case, the conservation equation yields the standard matter density evolution
\begin{equation*}
 \rho(a) = \frac{\rho_0}{a^3}
 \tag{21} \label{equ(21)}
\end{equation*}
Using the expression for the scale factor obtained previously, the matter density can be written as a function of the non-metricity scalar,
\begin{equation*}
 \rho(Q) = \frac{Q - 6H_0^2 \Omega_{\Lambda}}{16 \pi}.
 \tag{22} \label{equ(22)}
\end{equation*}
Substituting this relation into the differential equation \eqref{equ(16)}, we obtain
\begin{equation*}
2Q f_Q - f = -(16 \pi + 3 \lambda) \left(\frac{Q - 6H_0^2 \Omega_{\Lambda}}{16\pi } \right),
\tag{23} \label{equ(23)}
\end{equation*}
whose analytical solution is
\begin{equation*}
f(Q) = c_1\sqrt{Q} - (Q + 6H_0^2 \Omega_{\Lambda}) \frac{16 \pi + 3 \lambda}{16 \pi }
\tag{24} \label{equ(24)}
\end{equation*}
where $c_1$ is the integration constant. Thus, the reconstructed Lagrangian consists of a term proportional to $\sqrt{Q}$, arising from the homogeneous solution, and a linear term in $Q$ together with a constant contribution determined by the prescribed $\Lambda$CDM background.\\
Furthermore, in the limit $\lambda \to 0$, the reconstructed dust solution agrees with the corresponding reconstruction result in $f(Q)$ gravity \cite{gadbail2022reconstruction}.

\subsection{Reconstruction for a Perfect Fluid with \texorpdfstring{$\omega=-\frac13$}{w=-1/3}} \label{3.2}
We next consider a perfect fluid satisfying the equation of state $p=-\rho/3$. Imposing this value of the EoS parameter in the conservation equation, we obtain the corresponding matter density in terms of $a(t)$ as
\begin{equation*}
    \rho(a) = \frac{\rho_0}{a^2}.
    \tag{25} \label{equ(25)}
\end{equation*}
Using the previously derived expression for the scale factor in terms of the non-metricity scalar in equation \eqref{equ(20)}, the energy density becomes
\begin{equation*}
    \rho(Q) = \left(\frac{\rho_0}{(16 \pi)^2}\right)^{\frac{1}{3}} (Q - 6H_0^2 \Omega_{\Lambda})^{\frac{2}{3}}
    \tag{26} \label{equ(26)}
\end{equation*}
Substituting the above expression into the differential equation \eqref{equ(16)} yields
\begin{equation*}
    2Qf_Q - f = - \rho_0\left(16 \pi + \frac{10}{3} \lambda \right) \left(\frac{Q - 6H_0^2 \Omega_{\Lambda}}{16 \pi \rho_0} \right)^{\frac{2}{3}}.
    \tag{27} \label{equ(27)}
\end{equation*}
 In this scenario, the matter density in equation \eqref{equ(26)} is a non-integer power of $Q - 6H_0^2 \Omega_{\Lambda}$. Consequently, the corresponding first-order linear differential equation does not reduce to a polynomial particular solution, and its general solution is expressed in terms of the Gauss hypergeometric function. The resulting differential equation admits the analytical solution
\begin{equation*}
    f(Q) = c_1 \sqrt{Q} + \dfrac{\rho_0 (24 \pi + 5 \lambda)\left(\frac{Q - 6H_0^2 \Omega_{\Lambda}}{\rho_0} \right)^{\frac{2}{3}} \times ~ _2F_1 \left(-\frac{2}{3}, -\frac{1}{2}, \frac{1}{2}, \frac{Q}{6H_0^2 \Omega_{\Lambda}} \right)}{6 (2 \pi)^{\frac{2}{3}} \left(1 - \frac{Q}{6 H_0^2 \Omega_{\Lambda}} \right)^{\frac{2}{3}}},
    \tag{28} \label{equ(28)}
\end{equation*}
where $c_1$ is the integration constant and $_2F_1$ is a Hypergeometric function.\\
As a consistency check, we consider the limit $\lambda \to 0$, corresponding to a vanishing matter–geometry coupling. In this limit, the reconstruction equation reduces to the corresponding $f(Q)$ gravity equation, and the solution obtained above simplifies directly to the hypergeometric form of the $f(Q)$ reconstruction for $\omega = -\frac{1}{3}$ \cite{gadbail2022reconstruction}.

\subsection{Reconstruction for Nonisentropic Perfect Fluids} \label{3.3}
Finally, we investigate the reconstruction corresponding to a nonisentropic perfect fluid. In this case, the pressure is assumed to satisfy $p = \omega(a) \rho$.\\ We consider an example where the time-dependent barotropic index is given by
\begin{equation*}
    \omega(a) = \frac{2 \alpha - \beta a^3}{\alpha + \beta a^3},
    \tag{29} \label{equ(29)}
\end{equation*}
where $\alpha$ and $\beta$ are constants. The conservation equation can then be integrated to obtain the energy density as a function of the scale factor,
\begin{equation*}
    \rho(a) = \frac{(\alpha + \beta a^3)^3}{a^9}.
    \tag{30} \label{equ(30)}
\end{equation*}
Using the equation \eqref{equ(20)}, both the equation of state parameter and the matter density are expressed in terms of the non-metricity scalar,
\begin{equation*}
    \rho(Q) = \left(\frac{A(Q)}{16 \pi \rho_0} \right)^3,
    \tag{31} \label{equ(31)}
\end{equation*}
\begin{equation*}
    \omega(Q) = \frac{2 \alpha (Q - 6H_0^2 \Omega_{\Lambda}) - 16 \pi \beta \rho_0}{A(Q)},
    \tag{32} \label{equ(32)}
\end{equation*}
where $A(Q) = 16 \pi  \beta \rho_0 + \alpha (Q - 6H_0^2 \Omega_{\Lambda}).$\\
Substituting these expressions into the differential equation \eqref{equ(16)} leads to the corresponding reconstructed gravitational Lagrangian obtained as
\begin{equation*}
f(Q)=c_{1}\sqrt{Q}+\mu_{1}+\mu_{2}Q+\mu_{3}Q^{2}+\mu_{4}Q^{3},
\tag{33} \label{equ(33)}
\end{equation*}
where the coefficients $\mu_i$ depend on the model parameters and are given by
\begin{align*}
\mu_1 &=
\frac{
\left(8\pi \beta\rho_0-3H_0^2\alpha\Omega_{\Lambda}\right)^2
\left[
8\pi \beta\left(16\pi+(3+2\alpha)\lambda\right)\rho_0
-3H_0^2\alpha(16\pi+\lambda)\Omega_{\Lambda}
\right]
}{
512\,\pi^3\rho_0^3
}, \\[6pt]
\mu_2 &=
\frac{
\alpha\left(-8\pi \beta\rho_0+3H_0^2\alpha\Omega_{\Lambda}\right)
\left[
8\pi \beta\left(48\pi+(7+4\alpha)\lambda\right)\rho_0
-9H_0^2\alpha(16\pi+\lambda)\Omega_{\Lambda}
\right]
}{
1024\,\pi^3\rho_0^3
}, \\[6pt]
\mu_3 &=
\frac{
\alpha^2
\left[
-8\pi \beta\left(48\pi+(5+2\alpha)\lambda\right)\rho_0
+9H_0^2\alpha(16\pi+\lambda)\Omega_{\Lambda}
\right]
}{
6144\,\pi^3\rho_0^3
}, \\[6pt]
\mu_4 &=
-\frac{
\alpha^3(16\pi+\lambda)
}{
20480\,\pi^3\rho_0^3
}.
\end{align*}
As a consistency check, we consider the limit $\lambda \rightarrow 0$, where the matter--geometry coupling vanishes, and the theory reduces to pure $f(Q)$ gravity. Under geometrized units ($G = 1$), our matter-density normalization $\rho_0$ is related to the density $\rho_{0,\rm g}$ adopted in Reference \cite{gadbail2022reconstruction} by $\rho_{0,\rm g} = 8\pi \rho_0$. Taking $\lambda \rightarrow 0$ and applying this relation, our reconstructed Lagrangian coefficients reduce directly to those of Reference \cite{gadbail2022reconstruction}.

\section{Cosmological Reconstruction in Terms of the e-folding Parameter} \label{4}
In the previous section, the reconstruction procedure was carried out by expressing the scale factor and the matter density in terms of the non-metricity scalar $Q$. In this section, we present an alternative formulation of the reconstruction in terms of the e-folding parameter. This approach provides a convenient description of the cosmological evolution and allows the reconstruction equation to be expressed directly in terms of the e-folding variable.\\
We introduce the e-folding parameter as $ N=\ln a.$ Taking the present value of the scale factor to be $a_0=1$, the relation
between the scale factor and the cosmological redshift is $a=\frac{1}{1+z},$ and hence $ N=-\ln(1+z).$ 
Using the relation $1+z=e^{-N}$, the Hubble parameter given in the equation \eqref{equ(17)} can be expressed in terms of the e-folding parameter as
\begin{equation*}
    H^2(N) = H_0^2 [\Omega_{m_0} e^{-3N} + \Omega_{\Lambda}].
    \tag{34} \label{equ(34)}
\end{equation*}
Using the definition of the non-metricity scalar, $Q = 6H^2$, we obtain
\begin{equation*}
    Q(N) = 6H_0^2\left(\Omega_{m0}e^{-3N}+\Omega_{\Lambda}\right).
    \tag{35} \label{equ(35)}
\end{equation*}
Solving the above relation for $e^{-3N}$ gives
\begin{equation*}
    e^{-3N} = \frac{Q-6H_0^2\Omega_{\Lambda}}{6H_0^2\Omega_{m0}}.
    \tag{36} \label{equ(36)}
\end{equation*}
Consequently, the e-folding parameter can be expressed as a function of the non-metricity scalar in the form
\begin{equation*}
    N(Q) = -\frac{1}{3}\ln\left[\frac{Q-6H_0^2\Omega_{\Lambda}}
    {6H_0^2\Omega_{m0}} \right].
    \tag{37} \label{equ(37)}    
\end{equation*}
We now consider a perfect fluid with a constant equation of state
parameter $\omega$, such that $p=\omega\rho.$ Assuming that the matter sector satisfies the standard conservation equation,
\begin{equation*}
    \dot{\rho}+3H(\rho+p)=0,
    \tag{38} \label{equ(38)}
\end{equation*}
and using $p=\omega\rho$, we obtain
\begin{equation*}
    \dot{\rho}+3H(1+\omega)\rho=0.
    \tag{39} \label{equ(39)}
\end{equation*}
Since $ N=\ln a,$ we have $\frac{dN}{dt}=H.$ Therefore, the conservation equation can be written in terms of the e-folding parameter as
\begin{equation*}
    \frac{d\rho}{dN}
    +3(1+\omega)\rho=0.
    \tag{40} \label{equ(40)}
\end{equation*}
The solution of the above equation is
\begin{equation*}
    \rho(N) = \rho_0 e^{-3(1+\omega)N},
     \tag{41} \label{equ(41)}
\end{equation*}
where $\rho_0$ denotes the present-day matter density.
Using the relation obtained above for $e^{-3N}$, the matter density can
be expressed in terms of the non-metricity scalar as
\begin{equation*}
    \rho(Q) = \rho_0 \left[\frac{Q-6H_0^2\Omega_{\Lambda}}     {6H_0^2\Omega_{m0}}\right]^{1+\omega}.
    \tag{42} \label{equ(42)}
\end{equation*}
Using the relation between the present-day matter density and the matter
density parameter $ \Omega_{m0} =\frac{8\pi \rho_0}{3H_0^2}$, the above expression can be written as
\begin{equation*}
    \rho(Q) = \rho_0\left[\frac{Q-6H_0^2\Omega_{\Lambda}}{16\pi \rho_0}
    \right]^{1+\omega}.
    \tag{43} \label{equ(43)}
\end{equation*}
Substituting the above expression for the matter density into the
equation \eqref{equ(16)}, we obtain the first-order linear differential equation
\begin{equation*}
    2Qf_Q-f = -\left[16\pi+\lambda(3-\omega)\right]\rho_0\left[        \frac{Q-6H_0^2\Omega_{\Lambda}}{16\pi \rho_0}\right]^{1+\omega}.
    \tag{44}\label{equ(44)}
\end{equation*}
The above equation provides the reconstructed differential equation for
the gravitational Lagrangian $f(Q)$ corresponding to a constant
equation-of-state parameter $\omega$. Its solution can be obtained for
different choices of $\omega$, thereby providing the reconstructed
Lagrangian for the corresponding matter sources.\\
For example, for pressureless dust, $\omega=0$, the above equation reduces
to the dust reconstruction equation obtained in section (\ref{3.1}). Similarly, setting $\omega=-1/3$ reproduces the reconstruction equation corresponding to the perfect fluid considered in section (\ref{3.2}). Thus, the e-folding formulation provides a general representation that contains the
constant-$\omega$ cases considered previously as special cases.\\
The consistency between the results obtained using the scale-factor
formulation and the e-folding formulation demonstrate that the reconstruction procedure is independent of the choice of the cosmological variable used to parameterize the background evolution.

\section{Conclusion} \label{5}
A successful description of the observed late-time accelerated expansion of the Universe is provided by the $\Lambda$CDM model, although the nature of the dark sector and the underlying gravitational dynamics remain open questions. 
In this work, we considered the functional form $f(Q,T)=f(Q)+\lambda T$ and investigated the cosmological reconstruction of the gravitational Lagrangian in $f(Q,T)$ gravity. For the background evolution, we adopted the $\Lambda$CDM expansion history and derived the differential equation \eqref{equ(16)} governing the reconstruction of $f(Q)$. Further, we expressed the matter density as a function of the nonmetricity scalar $Q$, allowing the reconstruction for different matter sectors.

First, we obtained an analytic expression for a dust-like matter distribution \eqref{equ(24)}. We then analyzed a perfect fluid with $\omega=-1/3$; the resulting expression was a hypergeometric form of $f(Q)$ \eqref{equ(28)}. Finally, a nonisentropic perfect fluid was taken into account, which led to a more general polynomial form of the reconstructed gravitational Lagrangian that includes terms up to third order in $Q$, in addition to the homogeneous contribution proportional to $\sqrt{Q}$ \eqref{equ(33)}. These results demonstrate that the reconstructed form of the gravitational Lagrangian depends explicitly on both the matter content and the coupling parameter $\lambda$.

As a consistency check, we examined the limit $\lambda\rightarrow0$, in which the matter--geometry coupling vanishes, and the present framework reduces to the corresponding $f(Q)$ description. For all the matter configurations considered, the reconstructed solutions consistently recover the corresponding $f(Q)$ results. This provides an independent check on the reconstructed gravitational Lagrangians and on the role of the coupling parameter in the present formulation.

Furthermore, we reformulated the reconstruction in terms of the e-folding parameter, which provides a convenient description of the reconstructed quantities along the cosmological background. This formulation offers an alternative representation of the reconstructed solutions in terms of the cosmological evolution and can be useful for studying their behavior over different stages of the expansion history.

This analysis is restricted to the background dynamics of a flat FLRW spacetime. Since theories reproducing the same background expansion can remain degenerate at the level of homogeneous cosmological observables, an important direction for future investigation is the study of cosmological perturbations, structure growth, and gravitational-wave propagation in the reconstructed $f(Q, T)$ models \cite{najera2022cosmological,albuquerque2022designer}. Such analyses may provide observational signatures that distinguish the reconstructed models from $\Lambda$CDM and their corresponding $f(Q)$ limits. 

\section*{Acknowledgement}
Gauree Shanker is thankful to the Department of Science and Technology (DST), Government of India, for providing financial assistance under the FIST project (TPN-69301), vide the letter with Ref. No.: (SR/FST/MS-1/2021/104).

\section*{Data Availability Statement}
There are no new data associated with this article.

\bibliographystyle{unsrtnat}
\bibliography{REFERENCES}

\end{document}